%% file: LIM_ET_AL_2020.tex
\documentclass[journal]{IEEEtran}

\input{packages}
\input{macros}

\newcommand{\fix}[1] {\textcolor{black}{#1}}
\renewcommand{\add}[1] {\textcolor{black}{#1}}

\usepackage{xcolor} % for marking with different colors. 
\usepackage[normalem]{ulem} % for strike-throughs

\usepackage{amsmath} % for asterisk 

\begin{document}

\title{CycleGAN with a Blur Kernel for Deconvolution Microscopy: Optimal Transport Geometry}

\author{
Sungjun Lim\text{*},  Hyoungjun Park\text{*}, Sang-Eun Lee, Sunghoe Chang, Byeongsu Sim,
        and~Jong~Chul~Ye,~\IEEEmembership{Fellow,~IEEE}% <-this % stops a space
%\thanks{Copyright (c) 2017 IEEE. Personal use of this material is permitted. However, permission to use this material for any other purposes must be obtained from the IEEE by sending a request to pubs-permissions@ieee.org.}
\thanks{\text{*}: First co-authors with equal contributions to this work.
SL, HP and JCY are with the Department of Bio and Brain Engineering, Korea Advanced Institute of Science and Technology (KAIST), 
		Daejeon 34141, Republic of Korea (e-mail:\{jong.ye\}@kaist.ac.kr). BS and JCY are  with the Dept. of Mathematical Sciences, KAIST.
		SEL and SC are with the Department of Physiology \& Biomedical Sciences, Seoul National University College of Medicine, Seoul 03080, Republic of Korea. Part of this work was presented in \cite{lim2019}.}}% <-this % stops a space	

% make the title area
\maketitle

% As a general rule, do not put math, special symbols or citations
% in the abstract or keywords.
\begin{abstract}
Deconvolution microscopy has been extensively used to improve the resolution of the wide-field fluorescent microscopy, but the performance of classical approaches critically depends on the accuracy of a model  and optimization algorithms. Recently,  the convolutional neural network (CNN)  approaches have been 
studied as a fast and high performance alternative. Unfortunately, the CNN approaches usually require matched high resolution images for supervised training. In this paper,  we present a novel {\em unsupervised} cycle-consistent generative adversarial network (cycleGAN)  with a linear blur kernel,  which   can be used for both blind- and non-blind image
deconvolution. In contrast to the conventional cycleGAN approaches that require two deep generators, 
 the proposed cycleGAN approach needs  only a single deep generator and a linear blur kernel, which significantly improves the robustness and efficiency of network training. We show that the proposed architecture is indeed a dual formulation of an optimal transport problem that uses a special form of the penalized least squares cost as a  transport cost. 
Experimental results using simulated and real experimental data confirm the efficacy of the algorithm.
\end{abstract}

\begin{IEEEkeywords}
Deconvolution microscopy,  unsupervised learning,   generative adversarial network (GAN), cycle consistency, optimal transport, penalized least squares
\end{IEEEkeywords}

\IEEEpeerreviewmaketitle

\section{Introduction}

\IEEEPARstart{L}ight diffraction from a given optics limits the resolution of  images,
 creating blur and haze, which degrade fine details of the intracellular organelles. 
 Therefore,  deconvolution techniques are usually required to improve the resolution \cite{sarder2006deconvolution}.

 Mathematically, a blurred image $g(\rb),~\rb\in\Rd^3$ can be represented as a convolution between an unobserved object $x(\rb)$ and a 3D point-spread function (PSF) $h(\rb)$:
\begin{eqnarray}\label{eq:g}
g(\rb) &=&
\Hc x(\rb) \notag\\
&=&  h \ast x(\rb) = \int_{\mathbb{R}^{3}}h(\rb-\rb')x(\rb')d\rb',
\label{Eq1}
\end{eqnarray}
which is  also corrupted by noises:
\begin{eqnarray}\label{eq:forward}
y =&\Hc x+ w =& h\ast x + w \ , 
\end{eqnarray}
where $w$ is the measurement noise.

 It is well-known that a deconvolution microscopy problem, which obtains $x$ from the sensor measurement $y$, is ill-posed.
A standard strategy to address the ill-posed inverse problems is  the  penalized least squares approach  to stabilize the solution:
\begin{eqnarray}\label{eq:problem}
\min_{x} \|y -h \ast x\|^2+  R(x)
\end{eqnarray}
 where 
$R(x)$ is a model-based regularization (or penalty) function such as  $l_1$, total variation (TV), etc \cite{chaudhuri2014blind,sarder2006deconvolution,mcnally1999three}.

If the  PSF  $h(\rb)$ is not known,  which is often referred to as the blind deconvolution problem,
both the unknown PSF $h(\rb)$  and the image $x(\rb)$ must be estimated.
For example, the authors in \cite{you1996regularization,chan1998total} proposed a blind deconvolution method by solving a joint minimization problem to estimate both the unknown
blur kernel and  the image by adding additional  penalty (such as $l_1$ or TV) for the unknown blur kernel:
\begin{eqnarray}\label{eq:blind}
\min_{x,h} \|y - h \ast x\|^2+ R(x) + Q(h)
\end{eqnarray}
where $Q(h)$ is the regularization term for the blur kernel.
Eq.~\eqref{eq:blind} can be considered a general form of the penalized least squares formulation for deconvolution microscopy, 
since the non-blind case is considered a special case  with  $Q(h)=-\delta(h-h^*)$, where  $\delta(\cdot)$ is the delta function.

Recently, inspired by the success of convolutional neural network (CNN)
 for natural image deconvolution\cite{xu2014deep}, several researchers have employed CNN for optical microscopy deconvolution problems.
For example, Rivenson et al. \cite{rivenson2017deep} used deep neural networks to  enhance its spatial resolution for  a large field-of-view and depth-of-focus  optical microscopy. Nehme et al. \cite{nehme2018deep} used a deep convolutional neural network to obtain super resolution images from localization microscopy. 
 Weigert et al. \cite{weigert2017isotropic} proposed a CNN method which can recover isotropic resolution from anisotropic data. Zelger et al. \cite{zelger2018three} used a CNN method to localize particles in three dimensions from a single image.  In these works, the neural networks are usually trained in a supervised manner, necessitating high resolution reference images that are matched to each low resolution images.
 
Recently, the generative adversarial network (GAN) has attracted significant attention in inverse problems by providing a way to use unmatched data to train a deep neural network \cite{goodfellow2014generative}. In particular, the authors in \cite{arjovsky2017wasserstein} proposed so-called Wasserstein GAN, which  is closely related to the mathematical theory of optimal transport \cite{villani2008optimal,peyre2019computational}. 
For microscopy applications,
Nguyen et al. \cite{nguyen2018deep} developed a novel CNN framework with conditional generative adversarial network (cGAN) framework to reconstruct video sequences of dynamic live cells from Fourier Ptychographic Microscopy (FPM).  Pan et al. \cite{pan2020} added physics-based formulation that guides the estimation process of an image restoration task to a standard GAN structure. Unfortunately, these GAN approaches often generate  artificial features due to the mode collapsing, so a cycle-consistent adversarial network (cycleGAN) \cite{zhu2017unpaired} that imposes the one-to-one correspondence has  been extensively investigated for image reconstruction and super-resolution \cite{kang2019cycle,lu2017conditional, yuan2018}. In microscopy applications, Lee et al \cite{lee2019three} employed the standard cycleGAN for 3-D fluorescence microscopy. However,  cycleGAN is relatively difficult to train, since two deep neural network generators should be trained simultaneously.

In our recent paper\cite{sim2019optimal}, we demonstrated that
a general form of cycleGAN architecture can be derived from the Kantorovich's dual formulation of optimal transport (OT)  \cite{villani2008optimal,peyre2019computational} using a novel 
penalized least squares (PLS) transportation cost, where the physics-driven data consistency term is enforced as a regularization term in learning the transportation map.  
Inspired by the theory, one of the most important contribution of this paper is to show that for optical microscopy applications with the {unknown} blur model in \eqref{eq:forward}, the cycleGAN architecture with two deep generators may  not be necessary for unsupervised  deconvolution microscopy. Instead, a simple linear blur generator may be sufficient to generate low resolution images, thereby reducing the required number of CNN generators to only one. 
Moreover, for the non-blind case with a known PSF,  we show that the architecture can be further simplified so that
only a single pair of generator and discriminator is necessary.
These make the training much easier and more stable compared to the original cycleGAN approaches with two deep generators. We are aware that the optimal transport theory was also used for deconvolution problem \cite{schmitz2018wasserstein}. However, the main idea of \cite{schmitz2018wasserstein} is to design a dictionary for the point spread function using the optimal transport theory, whereas our method tries to develop a cycleGAN architecture to estimate the image directly.

This paper is composed as follows. In 
Section~\ref{sec:theory}, we propose a cycleGAN architecture with a linear blur kernel as a dual
formulation of the optimal transport problem with the penalized least squares. The network implementation
 is detailed in Section~\ref{sec:network}.
The experimental methods and results are provided 
in Section~\ref{sec:methods} and Section~\ref{sec:results}, respectively, which are followed by the conclusion in Section~\ref{sec:conclusion}.

\section{Theory}\label{sec:theory}

\subsection{Optimal Transport Driven CycleGAN}
\begin{figure}[!hbt]
\centering\includegraphics[width=0.45\textwidth]{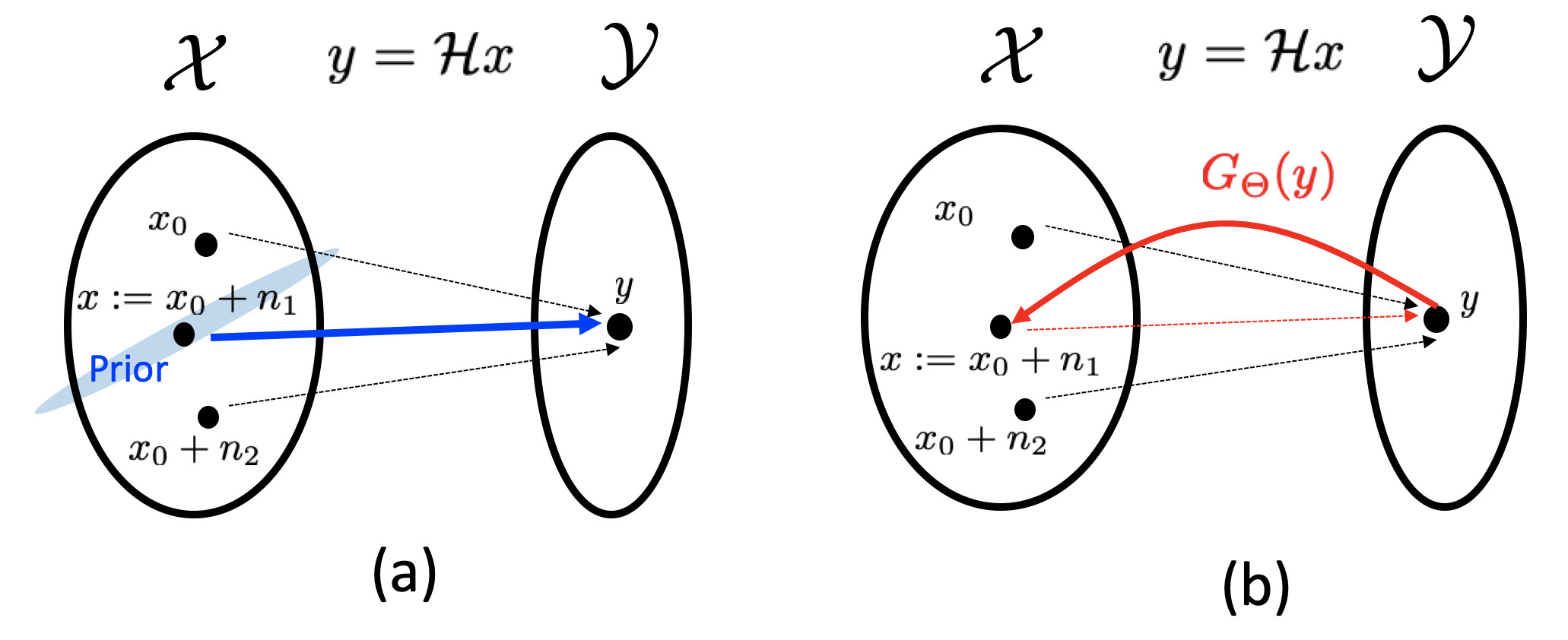}
\caption{Two strategies for resolving ambiguities in the feasible solutions in an ill-posed inverse problem. (a) Classical PLS approach using a close form prior distribution,
and (b) our PLS approach using an inverse mapping to define a prior. }
\label{fig:concept}
\end{figure}

An exemplary geometric view of the deconvolution problem from the forward model \eqref{eq:forward} is shown in Fig.~\ref{fig:concept},
where the mapping $\Hc:\Xc\mapsto\Yc$ converts high resolution image manifold $\Xc$  
to the measurement manifold $\Yc$ for the blurry images.  Let the probability measures supported on the $\Xc$ and $\Yc$ spaces are denoted by
$\mu$ and $\nu$, respectively.

{
The main technical difficulty of this deconvolution problem is its ill-posedness.
For example,
even for  a non-blind problem where the blur kernel $h$ is known, the main difficulty lies in  the existence of null space of the forward operator $\Hc$.
For example, suppose that  $n_1$ and  $n_2$ are both from the null space of $\Hc$, i.e. $n_1,n_2 \in \Nc(\Hc)$,
where as $x_0 \in \Rc(\Hc^\top)$
such that
$ \Hc(x_0+n_1) = \Hc(x_0+n_2)$.
Here, $\Nc(\cdot)$ and $\Rc(\cdot)$ denote the null space and the range space, respectively; and the superscript $^\top$ denotes the adjoint operation.
The assumption essentially means that any linear method will only be able to recover the component $x_0$.  
This in practice for deconvolution microscopy poses issues: because not all features in the high resolution
 space are  in $\Rc(\Hc^\top)$, many are in $\Nc(\Hc)$.
 }

{To recover the missing features in the null space, the classical PLS methods use a prior knowledge in terms of regularization.
 In particular, the prior knowledge restricts the feasible solution set of the optimization problems and reduces the ill-posedness of 
 the solution as shown in Fig.~\ref{fig:concept}(a).  Accordingly, a solution $x_0+n_1$ which is not in $\Rc(\Hc^\top)$ could be recovered.
  Therefore, the  key engineering issue here lies in the
 design of the regularization terms such that the resulting feasible solution sets contain the true high resolution features.}

{On the other hand, in the recent GAN-based  super-resolution approaches \cite{nguyen2018deep},
a generator $G_\Theta:\Yc \mapsto \Xc$ with the network 
parameter $\Theta$ and the input $y\in \Yc$ is designed to generate a matched high resolution image $x\in \Xc$
from a blurry image $y\in \Yc$.  As a result, if the null space component exists in the true $x\in \Xc$, one role of $G_\Theta(y)$ is to predict this component. 
Unfortunately, one of the main weaknesses of these GAN approaches is that the generated image $x$ is not guaranteed
to satisfy the data consistency, i.e. it is not clear whether the measurement can be reproduced when the generated image
is used as an input for the blur kernel.}

{This observation led us to propose a
novel PLS cost function \cite{sim2019optimal}:
\begin{eqnarray}\label{eq:cost0}
c(x,y;\Theta,\Hc)=\|y-\Hc x\|+ \| G_\Theta(y) -x \|
\end{eqnarray}
The new PLS cost function has many important  advantages over the GAN approaches. In particular,
if  the global minimizer is achieved, i.e. $c(x,y;\Theta,\Hc)=0$,  then  we have
\begin{eqnarray}\label{eq:primal}
 y=\Hc x,\quad x=G_{\Theta}(y)
\end{eqnarray}
Therefore, $G_{\Theta}$ can be an inverse of the forward operator $\Hc$, which is the ultimate goal
in the inverse problem. 
The geometric view of this situation is illustrated in Fig.~\ref{fig:concept}(b).
Specifically, in contrast to the classical PLS approaches that uses the prior knowledge to restrict the feasible set (see Fig.~\ref{fig:concept}(a)),
our PLS cost directly imposes the inversion path as a regularization term so that it generates the physically meaningful measurement.
Even when the global minimum is not achieved,
the new PLS cost \eqref{eq:cost0} employs the data consistency to regularize the
generator training so that the trained generator can generate physically consistent measurement.
}

Given these advantages of the new PLS cost \eqref{eq:cost0}, the critical question is how to determine the optimal parameter $\Theta$. This is where the optimal transport theory \cite{villani2008optimal,peyre2019computational} comes into play.

Specifically, since we do not have a matched low- and high-resolution data,
 the neural network parameter $\Theta$ should be estimated by minimizing the new PLS cost in \eqref{eq:cost0} for all 
combinations of $x \in \Xc$ and $y \in \Yc$ with respect to the joint measure 
$\pi (x,y)$.  According to the optimal transport theory \cite{villani2008optimal,peyre2019computational},
this can be addressed by defining the following average cost:
\begin{eqnarray}
\Kd(\Theta, \Hc) := \min_\pi \int_{\Xc \times \Yc} c(x,y;\Theta, \Hc)d\pi(x,y)
\label{eq:org}
\end{eqnarray}
where the minimum is taken over all joint distribution whose marginal distributions with 
respect to $\Xc$ and $\Yc$ are $\mu$ and $\nu$, respectively. 
In fact, the joint distribution $\pi$ which  minimizes \eqref{eq:org} is called the {\em optimal
 transportation map} \cite{villani2008optimal,peyre2019computational}.
 From the optimal transport theory perspective, the remaining problem
 is to estimate the unknown parameter $\Hc$ and $\Theta$ so that
 the average transportation cost in \eqref{eq:org} can be minimized.
 This is indeed a  {\em primal} form of the optimal transport \cite{villani2008optimal,peyre2019computational}.

Another important contribution of our companion paper  \cite{sim2019optimal} is that
the primal OT problem can be equivalently represented by the following {\em dual} OT problem using the Kantorovich dual formulation \cite{villani2008optimal,peyre2019computational}:
\begin{align}
\min_{\Theta, \Hc} \Kd(\Theta, \Hc) = \min_{\Theta, \Hc} \max_{\Phi, \Xi} 
\ell(\Theta, \Hc; \Phi, \Xi)
\end{align}
where 
\begin{eqnarray}
\ell(\Theta, \Hc;\Phi, \Xi) = \gamma \ell_{cycle}(\Theta, \Hc) + \ell_{WGAN}(\Theta, \Hc;\Phi, \Xi)
\end{eqnarray}
where $\gamma$ is an appropriate hyperparameter, 
$\ell_{cycle}$ denotes the cycle consistency loss, and $\ell_{WGAN}$ is the Wasserstein 
GAN \cite{arjovsky2017wasserstein} loss. 
More specifically, $\ell_{cycle}$ is given by \cite{sim2019optimal}:
\begin{eqnarray}
\begin{split}
\ell_{cycle}(\Theta, \Hc) =  \int_\Xc \|x - G_\Theta(\Hc x)\|d\mu(x) \\
+  \int_\Yc \|y - \Hc G_\Theta(y)\| d\nu(y)
\end{split}
\label{eq:cycle loss}
\end{eqnarray}
and  $\ell_{WGAN}$ is given by \cite{sim2019optimal}:
\begin{eqnarray}
\begin{split}
&\ell_{WGAN}(\Theta, \Hc;\Phi, \Xi) \\
&= \left(\int_\Xc \varphi_\Phi(x)d\mu(x) - \int_\Yc \varphi_\Phi(G_\Theta(y))d\nu(y)\right) \\
&+ (\left(\int_\Yc \psi_\Xi(y)d\nu(y) - \int_\Xc \psi_\Xi(\Hc x)d\mu(x)\right)
\end{split}
\label{eq:WGAN loss}
\end{eqnarray} 
where Kantorovich potentials $\varphi_\Phi$ and $\psi_\Xi$ should be 1-Lipschitz functions.
In GAN literature,  $\varphi_\Phi$ and $\psi_\Xi$ are often called the
Wasserstein GAN discriminators.
In fact, this formulation is similar to the standard cycleGAN \cite{zhu2017unpaired} except that the one of the generators
is now replaced with the physics-driven generator $\Hc x$, which imposes the
data consistency.

\subsection{Derivation of CycleGAN with a Linear Blur Kernel}

\subsubsection{Blind Deconvolution Cases}
To apply this idea for our optical microscopy deconvolution problem, only
minor modification is required. More specifically,  our PLS cost is now defined as
\begin{eqnarray}\label{eq:ourcost}
c(x,y;\Theta,h)=\|y-h\ast x\|+ \| G_\Theta(y) -x \| %+ Q(h)
\end{eqnarray}
This leads to the following dual OT formulation:
\begin{align}\label{eq:opt}
\min\limits_{\Theta,h}\max\limits_{\Phi,\Xi}\ell(\Theta,h;\Phi,\Xi)
\end{align}
where
\begin{eqnarray*}
\ell(\Theta,h;\Phi,\Xi) =\gamma \ell_{cycle}(\Theta,h)+ Q(h)+ \ell_{GAN}(\Theta,h;\Phi,\Xi)
\end{eqnarray*}
where the cycle-consistency loss
$\ell_{cycle}$ is given by:
\begin{eqnarray}\label{eq:ellcycle}
\begin{split}
\ell_{cycle}(\Theta, h) =  \int_\Xc \|x - G_\Theta(h\ast x)\|d\mu(x) \\
+  \int_\Yc \|y - h\ast G_\Theta(y)\| d\nu(y)
\end{split}
\end{eqnarray}
and $\ell_{WGAN}$ is the same as \eqref{eq:WGAN loss}  by replacing $\Hc x$ with $h\ast x$.

\subsubsection{Non-Blind Deconvolution Cases}
For the case of non-blind deconvolution, the PSF kernel $h$ is known, so it should not be estimated.
Therefore, competition with the associated discriminator $\psi_\Xi$ is not necessary.
This leads to a simplified optimization problem:
\begin{align}\label{eq:opt2}
\min\limits_{\Theta}\max\limits_{\Phi}\ell(\Theta;\Phi)
\end{align}
where
\begin{eqnarray*}
\ell(\Theta;\Phi) =\gamma \ell_{cycle}(\Theta)+  \ell_{GAN}(\Theta;\Phi,)
\end{eqnarray*}
where 
$\ell_{cycle}$ is simplified from \eqref{eq:ellcycle} to the following form:
\begin{eqnarray}
\begin{split}
\ell_{cycle}(\Theta) =  \int_\Xc \|x - G_\Theta(h\ast x)\|d\mu(x) \\
+  \int_\Yc \|y - h\ast G_\Theta(y)\| d\nu(y)
\end{split}
\end{eqnarray}
and
$\ell_{WGAN}$ is simplified  as
\begin{align}\label{eq:ourWGANnonblind}
&\ell_{WGAN}(\Theta;\Phi) \\
&= \left(\int_\Xc \varphi_\Phi(x)d\mu(x) - \int_\Yc \varphi_\Phi(G_\Theta(y))d\nu(y)\right) %\notag \\
%&+\eta_x \int_{{\Xc}}(\|\nabla_{\tilde x}\varphi_\Phi(x)\|_2 - 1)^2d\mu(x) \notag
\end{align}
where $\varphi_\Phi$ is again 1-Lipschitz function.

\begin{figure}[hbt!]
\centering\includegraphics[width=0.4\textwidth]{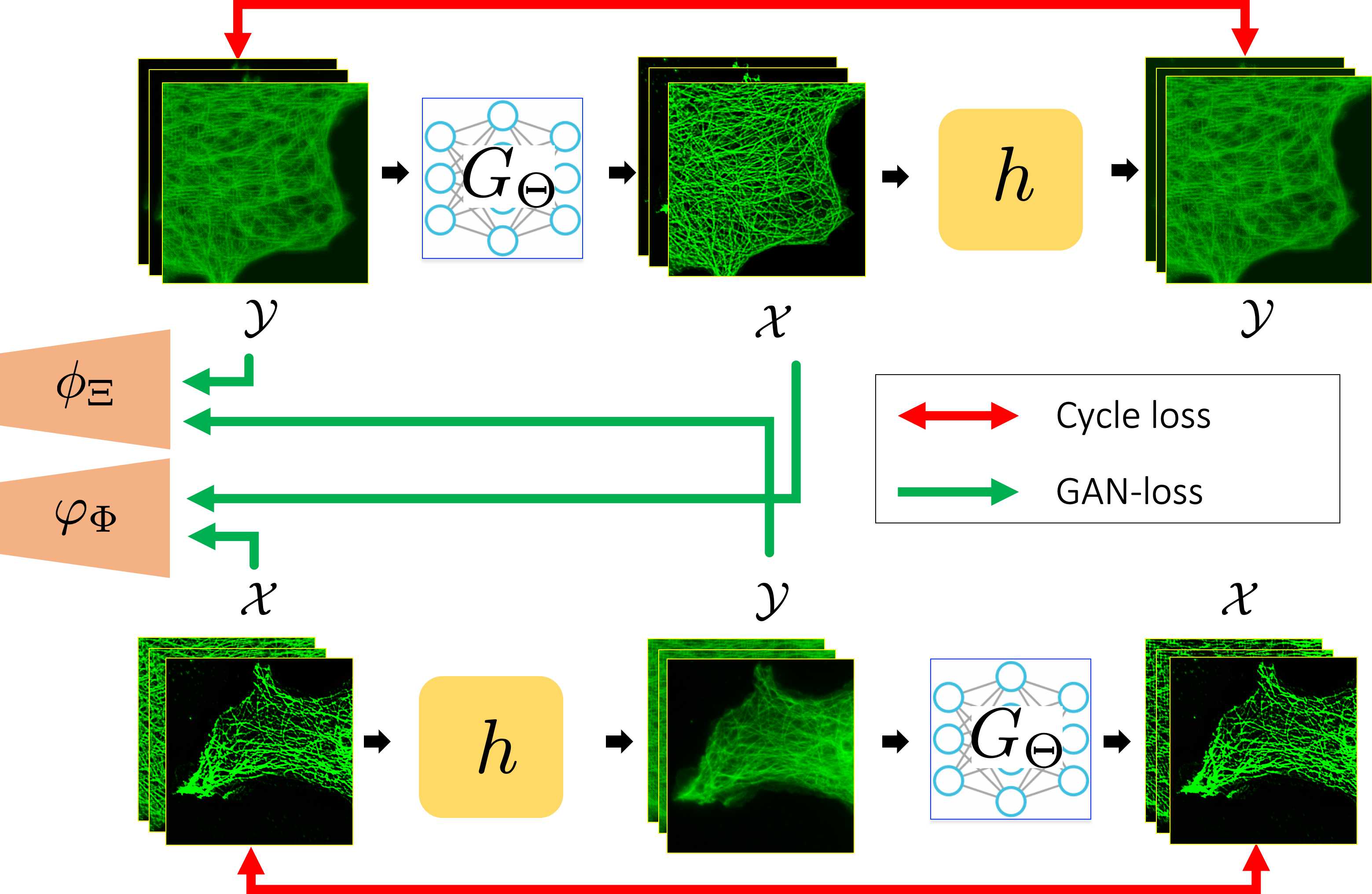}
\vspace*{0.5cm}
\centerline{(a)}
\centering\includegraphics[width=0.4\textwidth]{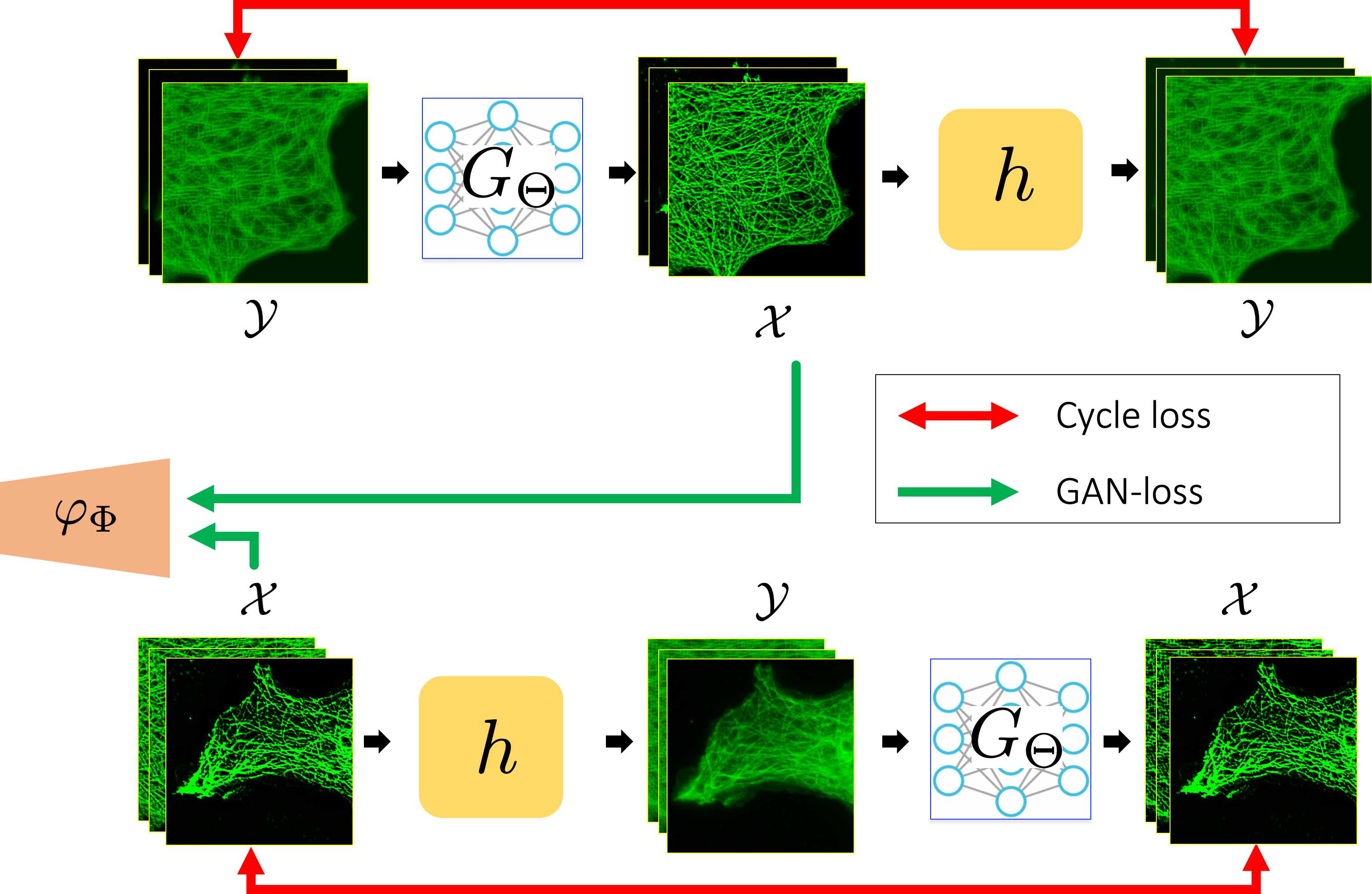}
\centerline{(b)}
\caption{Proposed cycleGAN architectures with a blur kernel for deconvolution microscopy for (a) blind cases and (b) non-blind case, respectively. Here,
$G_\Theta$ denotes the CNN-based low-resolution to high-resolution generator.  The blur generator is composed
of a linear blur kernel $h$. In addition, $\varphi_\Phi$ and $\phi_\Xi$ refer to the
CNN-based discriminators for the high resolution image domain $\Xc$ and the low-resolution image domain $\Yc$, respectively. 
 }
\label{fig:cycleGAN}
\end{figure}

\subsection{1-Lipschitz constraint}
For the implementation of $\ell_{WGAN}$,  care should be taken  to ensure that the Kantorovich potential  becomes 1-Lipschitz.
For example, in WGAN with the gradient penalty (WGAN-GP),
the gradient of the Kantorovich potential is constrained to be 1 \cite{gulrajani2017improved}.
Specifically, 
$\ell_{WGAN}$ is modified as
\begin{align}\label{eq:ourWGANnonblind2}
&\ell_{WGAN}(\Theta;\Phi) \\
&= \left(\int_\Xc \varphi_\Phi(x)d\mu(x) - \int_\Yc \varphi_\Phi(G_\Theta(y))d\nu(y)\right) \notag \\
&-\eta \int_{{\Xc}}(\|\nabla_{\tilde x}\varphi_\Phi(x)\|_2 - 1)^2d\mu(x) \notag
\end{align}
where 
$\eta>0$ is the regularization parameters to impose 1-Lipschitz property for the discriminators, and 
  $\tilde x=\alpha x+(1 - \alpha)G_\Theta(y)$ with $\alpha$ 
being random variables from the uniform distribution between $[0,1]$ \cite{gulrajani2017improved}. 
This approach is somewhat complicated due to the random sampling in calculating the gradient.

Another  approach is imposing 
a constraint on the magnitude of Kantorovich potential. This idea is  related to the `weight clipping' in the original Wasserstein GAN \cite{arjovsky2017wasserstein}.
More specifically, \eqref{eq:ourWGANnonblind} is implemented by constraining the magnitude of
$\varphi_\Phi(x)$ and  $\varphi_\Phi(G_\Theta(y))$ with respect to the discriminator parameter $\Phi$.
Since $\ell_{WGAN}(\Theta;\Phi)$ in  \eqref{eq:ourWGANnonblind} is given in the form of maximization with respect to $\Phi$,
the resulting regularized version of WGAN loss is given by
\begin{align}%\label{eq:ourLSnonblind}
&\ell_{WGAN}(\Theta;\Phi) \\
&= \left(\int_\Xc \varphi_\Phi(x)d\mu(x) - \int_\Yc \varphi_\Phi(G_\Theta(y))d\nu(y)\right) \notag \\
&-\eta \left( \int_{{\Xc}} \varphi_\Phi^2(x)d\mu(x) + \int_\Yc \varphi_\Phi^2(G_\Theta(y))d\nu(y) \right)
\end{align}
for some regularization parameter $\eta>0$. If we choose $\eta=\frac{1}{2}$, then we have
\begin{align}\label{eq:ourLSnonblind}
&\ell_{WGAN}(\Theta;\Phi)=   \\
&-\frac{1}{2} \left(\int_\Xc  (\varphi_\Phi(x)-1)^2d\mu(x) - \int_\Yc \left(\varphi_\Phi(G_\Theta(y))+1\right)^2d\nu(y)\right) \notag
%&-\eta \left( \int_{{\Xc}} \varphi_\Phi^2(x)d\mu(x) + \int_\Yc \varphi_\Phi^2(G_\Theta(y))d\nu(y) \right)
\end{align}
which is indeed the same as 
the least square GAN (LS-GAN) loss \cite{mao2017least}.
We found that LS-GAN loss is more efficient and stable than WGAN-GP for our deconvolution problems.

\section{Network Implementation}
\label{sec:network}
The  optimization problem \eqref{eq:opt} and \eqref{eq:opt2} can be implemented using 
a novel cycleGAN architecture with a linear blur kernel as shown in Fig.~\ref{fig:cycleGAN}(a) and (b), respectively.
Note that for the case of non-blind deconvolution shown in in Fig.~\ref{fig:cycleGAN}(b), only a single discriminator is used.
Here, we describe each block in more detail.

\subsection{Generator architecture}

\begin{figure}[!hbt]
\centering\includegraphics[width=0.4\textwidth]{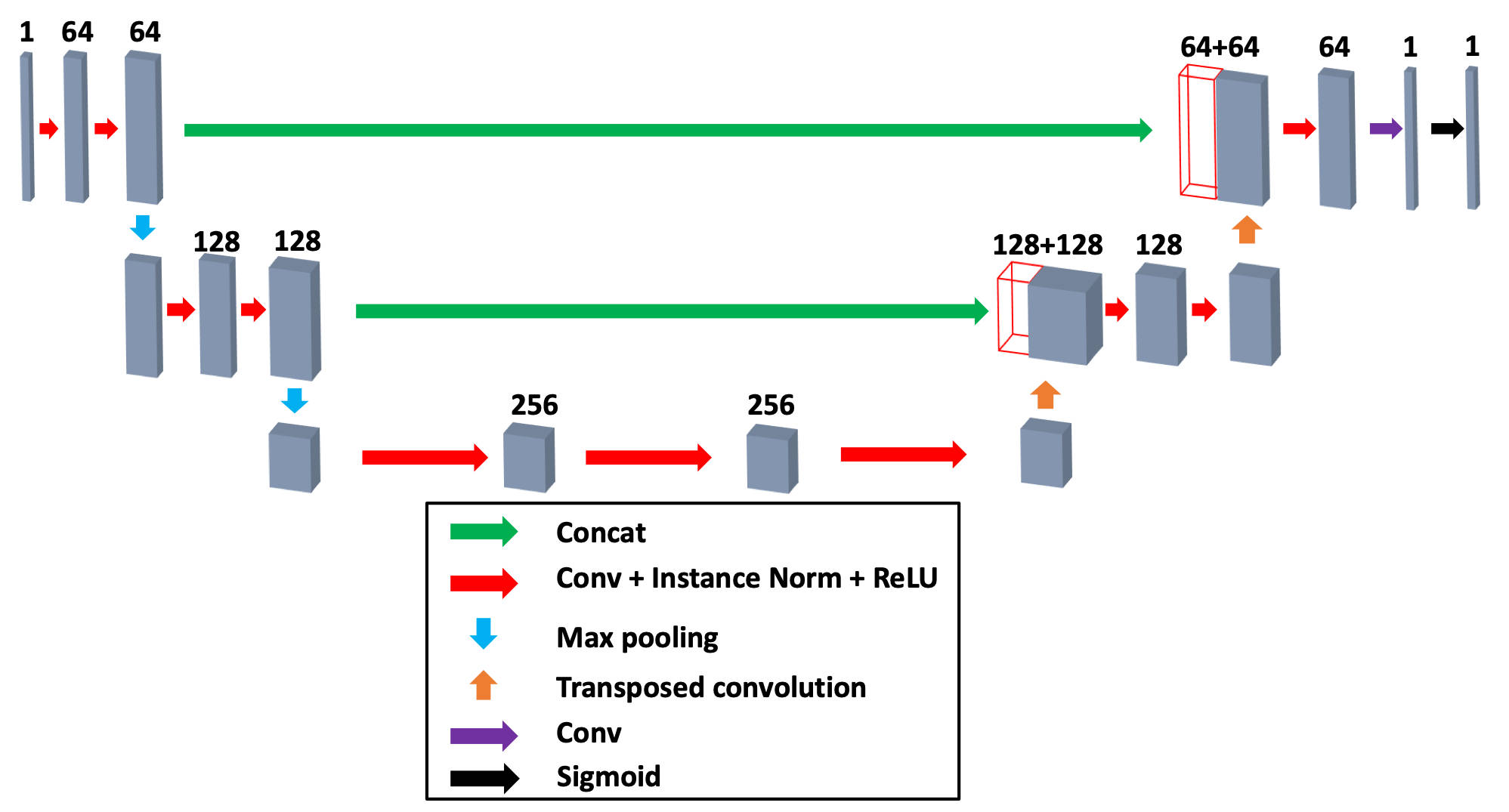}
\caption{A modified 3D U-net architecture for our high-resolution image generator.  
}
\label{fig:Unetwork}
\end{figure}

The network architecture of the high resolution image  generator $G_{\Theta}$ from the low-resolution image is a modified 3D-Unet \cite{cciccek20163d} as shown in Fig. \ref{fig:Unetwork}.
Our U-net structure consists of contracting and expanding paths. The contracting path consists of the repetition of the following blocks: {3D conv}- {Instance Normalization} \cite{ulyanov2016instance}- {ReLU}.
Here, the generator has  symmetric configuration  so that
both encoder and decoder have the same number of layers, i.e. $\kappa=7$.
Throughout the network, the convolutional kernel dimension is $3\times3\times3$.
There exists a pooling layer and skipped connection  for every other convolution operations.
To enhance the image contrast, we add an additional sigmoid layer at the end of U-Net.

\begin{figure}[h!]
\centering\includegraphics[width=0.4\textwidth]{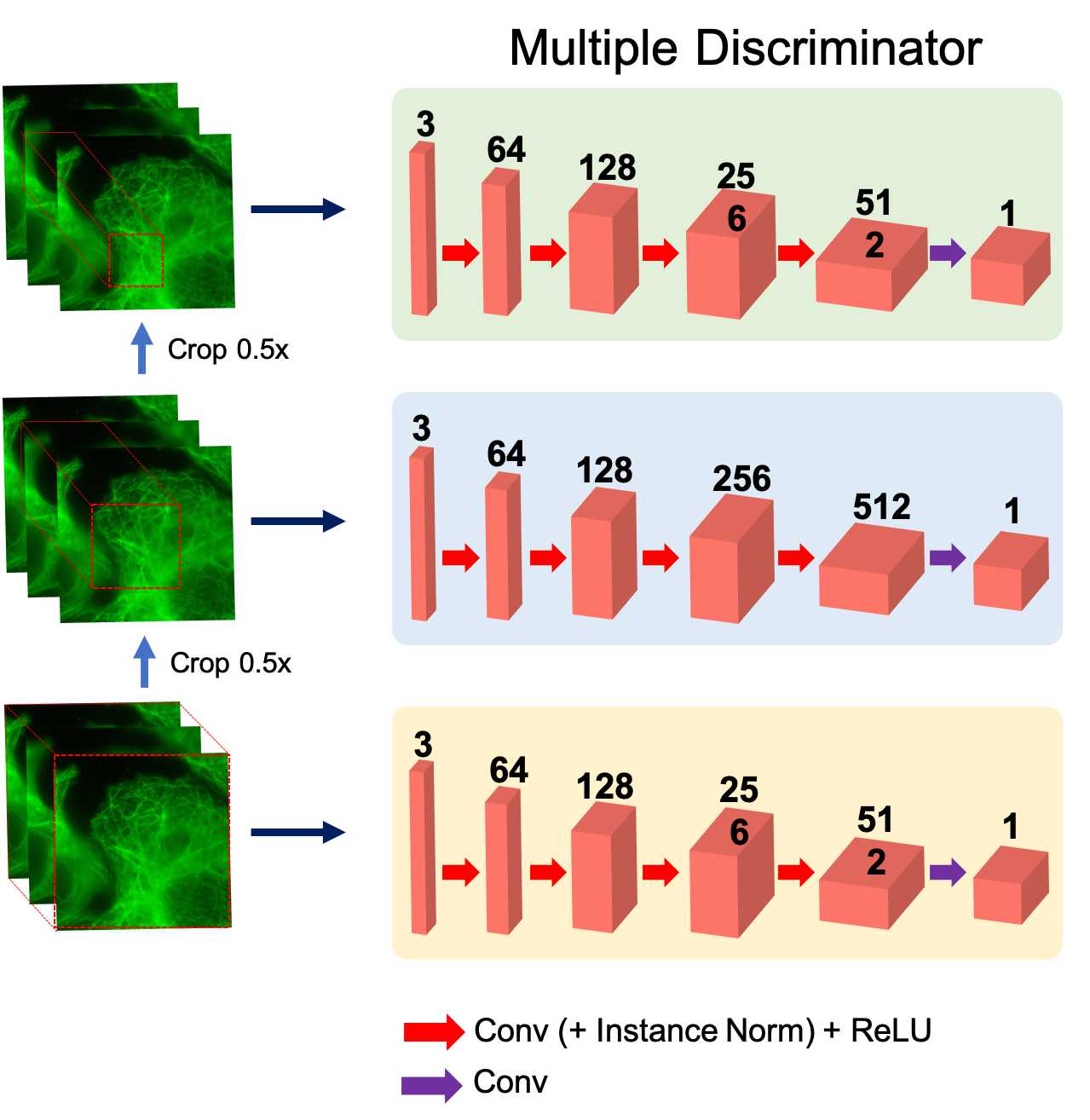}
\caption{Multi-PatchGANs discriminator  architecture.}
\label{fig:multiPatchGANArchitec}
\end{figure}

On the other hand, the low-resolution image generator  from high resolution input
is based on a single 3D  convolution layer that  models a 3D blurring kernel. 
The size of the 3D PSF modeling layer is chosen depending on the problem set by considering their approximate PSF sizes.  \ {In this paper, the size of the 3D PSF {kernel} is set to {$31\times 31\times 31$ } pixels  for the simulation study. {For real-world experiments, it is set to $20\times 20\times 20$ pixels.} }
%As for the penalty for the filter for blind deconvolution,
% we use the $l_1$ loss:
%\begin{eqnarray}\label{eq:l1} 
% Q(h)= \tau \|h\|_1,
% \end{eqnarray} 
% for some positive parameter $\tau$.

\subsection{Discriminator architecture }

As for the discriminators, we follow the original cycleGAN that uses multi-PatchGANs (mPGANs) \cite {isola2017image}, 
where each discriminator has input patches with different sizes used. 
As shown in Fig \ref{fig:multiPatchGANArchitec},  it consists of three independent discriminators. Each discriminator takes patches at different sizes: original, and  half, and quarter size patches. 

In place of W-GAN loss, we used the  least square GAN  (LS-GAN) loss \cite{mao2017least} {since it led to better performance than WGAN-GP \cite{gulrajani2017improved}.}  
To reduce the model oscillation \cite{goodfellow2014generative},  the discriminator in both models uses a history of generated volumes from a frame buffer containing 50 previously generated volumes, as described in \cite{shrivastava2017learning}.

\begin{figure}[!htb]
\begin{center}
\includegraphics[width=\linewidth]{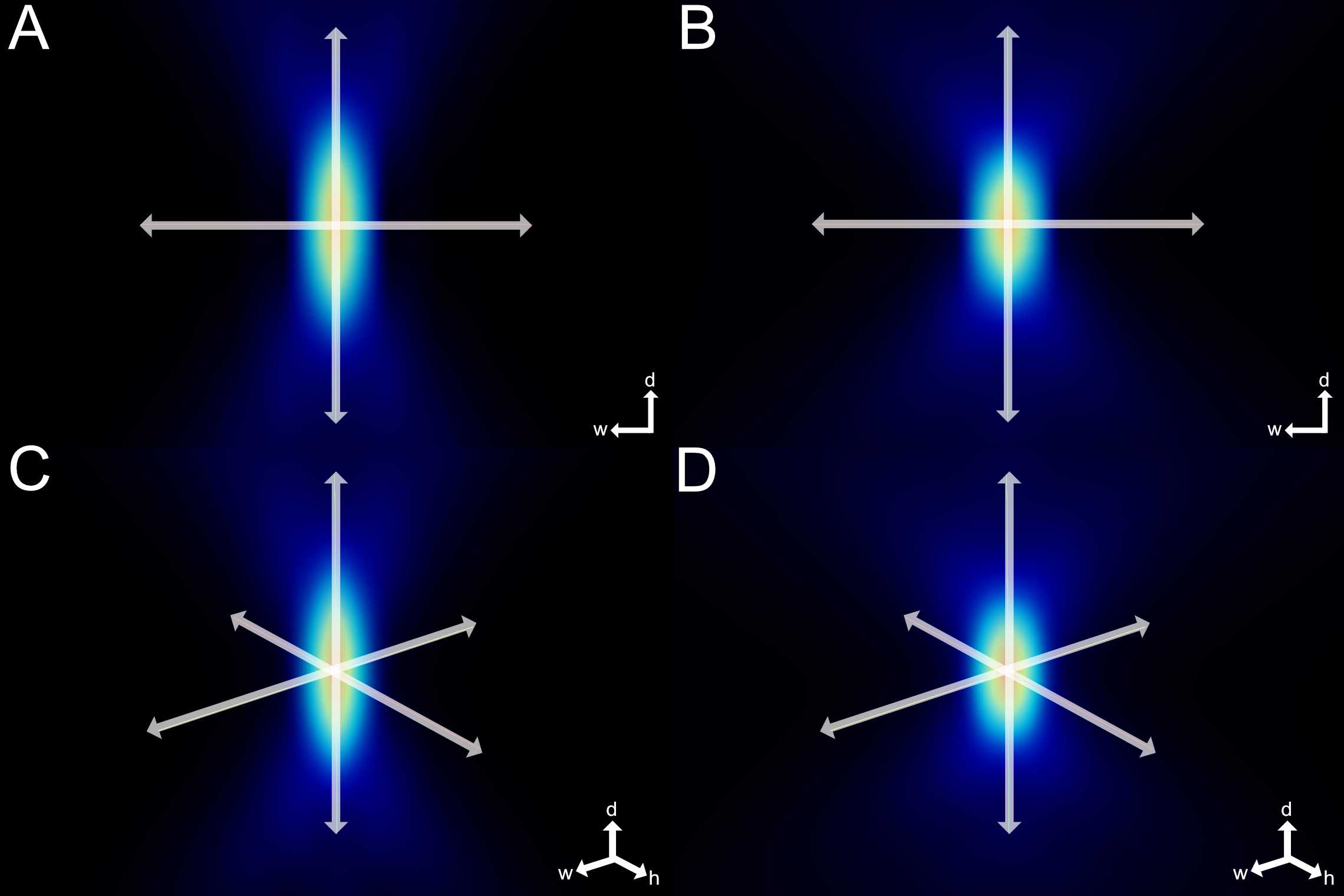}
\end{center}
\caption{Visualization of the PSF kernels: A, C are a 3D PSF image volume by the Born {\&} Wolf model and B, D are by the Richards {\&} Wolf model. (d, h, w) are  depth, height, width of an image volume. }
\label{fig:PSFkernels}
\end{figure}

\section{Method}
\label{sec:methods}

\subsection{Simulation Study}

\subsubsection{Processing of data-set}
For simulation studies with the ground-truth data, we used the synthetic microtubule network data set \cite {sage2017deconvolutionlab2}  to train, validate, {and test} our model. Specifically, from the ground-truth high resolution synthetic microtubule images,
we generate blurred images by convolving with a model PSF. { We used the PSF Generator tool implemented by Kirshner et al. \cite{kirshner2013} to simulate such PSF models.}

Specifically, the numerical PSF is set as the Born {\&} Wolf model \cite{born2013principles}, which describes the scalar-based diffraction that occurs in the microscope when the particle is in focus while the imaging plane may not be in focus. This is formulated as:

$$h(r_x,r_y,r_z)=$$
\begin{align}\label{eq:psf}
\Bigl| C \int_{0}^{1} J_{0} \Bigl[k \frac{NA}{n_{i}}\rho \sqrt{r_x^2+r_y^2} \Bigr] e^{ -\frac{1}{2} j k \rho^{2} r_z \left(\frac{NA}{n_{i}}\right)^{2}} \rho d\rho \Bigr|^{2}
\end{align}
where $C$ is a normalizing constant, $k = 2\pi / \lambda$ is the wave number of emitted light, $\lambda$ is the wavelength, $NA$ is the numerical aperture,  and $n_{i}$ is the refractive index of  immersion layer. 
For all simulations, we use $NA=1.4$, %the lateral  pitch was set to  100nm,  
and $n_i=1.5$. In our baseline model for training, $\lambda$ is set at $500nm$ and the voxel size is at $100nm$. The PSF models all are designed to have a $31\times31\times31$ kernel size. After convolving with a PSF kernel,  blurred image volumes were added \add{consecutively} with Gaussian noise and Poisson noise at multiple \fix{average} SNR (Signal-to-Noise) levels. \add{Here, the \fix{average} SNR is calculated with respect to the noise component, which is given by the difference between the resulting noisy image and the blurred image.}

In the inference step, we also tested for comparison on images that are blurred by a different PSF model, the Richards {\&} Wolf model \cite{richards_1959}, which describes the vectorial-based diffraction in the microscope and is set with the same wavelength and voxel size. We omit its formulation here. The kernels for the Born {\&} Wolf model and the Richards {\&} Wolf model are visualized in Figure \ref{fig:PSFkernels}.

\begin{figure*}[hbt!]
\centering\includegraphics[width=0.85\textwidth]{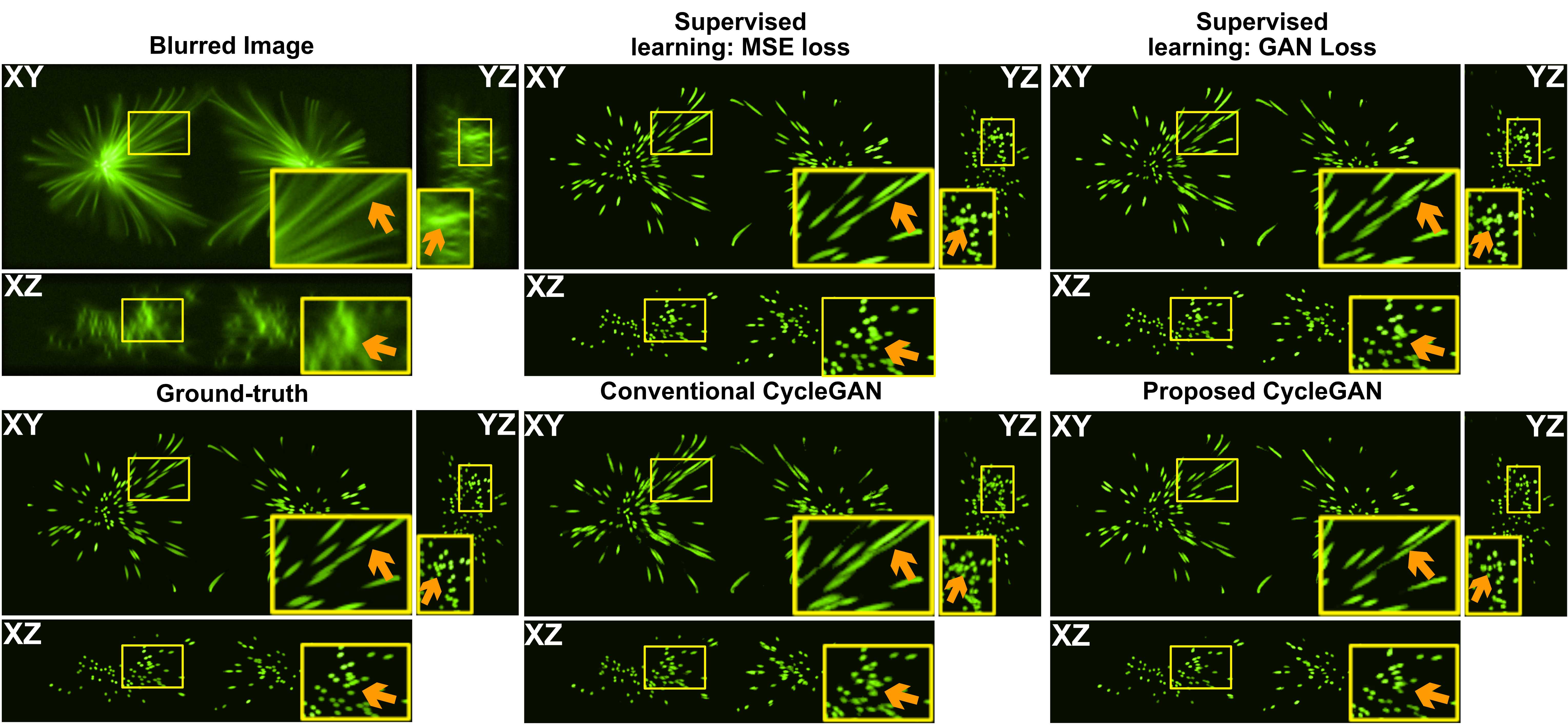}
\caption{Comparison of proposed methods with state-of-the-art deep learning methods for deconvolution. 
Synthetically generated microtubule network images were convolved with a Born {\&} Wolf PSF($\lambda=500$nm) and added with Gaussian and Poisson noise \fix{(peak SNR=24dB)} of 20dB to generate a realistically blurred image. Note that a cycleGAN architecture does not require paired data for training. The ROIs (marked yellow) show the area for the enlarged parts. \add{Orange arrows, as reference points, indicate the same location in each image.} XY, YZ, and XZ are 2D slicing planes of a 3D image volume. }
\label{fig:sota_comparison}
\end{figure*}

\begin{figure*}[hbt!]
\centering\includegraphics[width=0.85\textwidth]{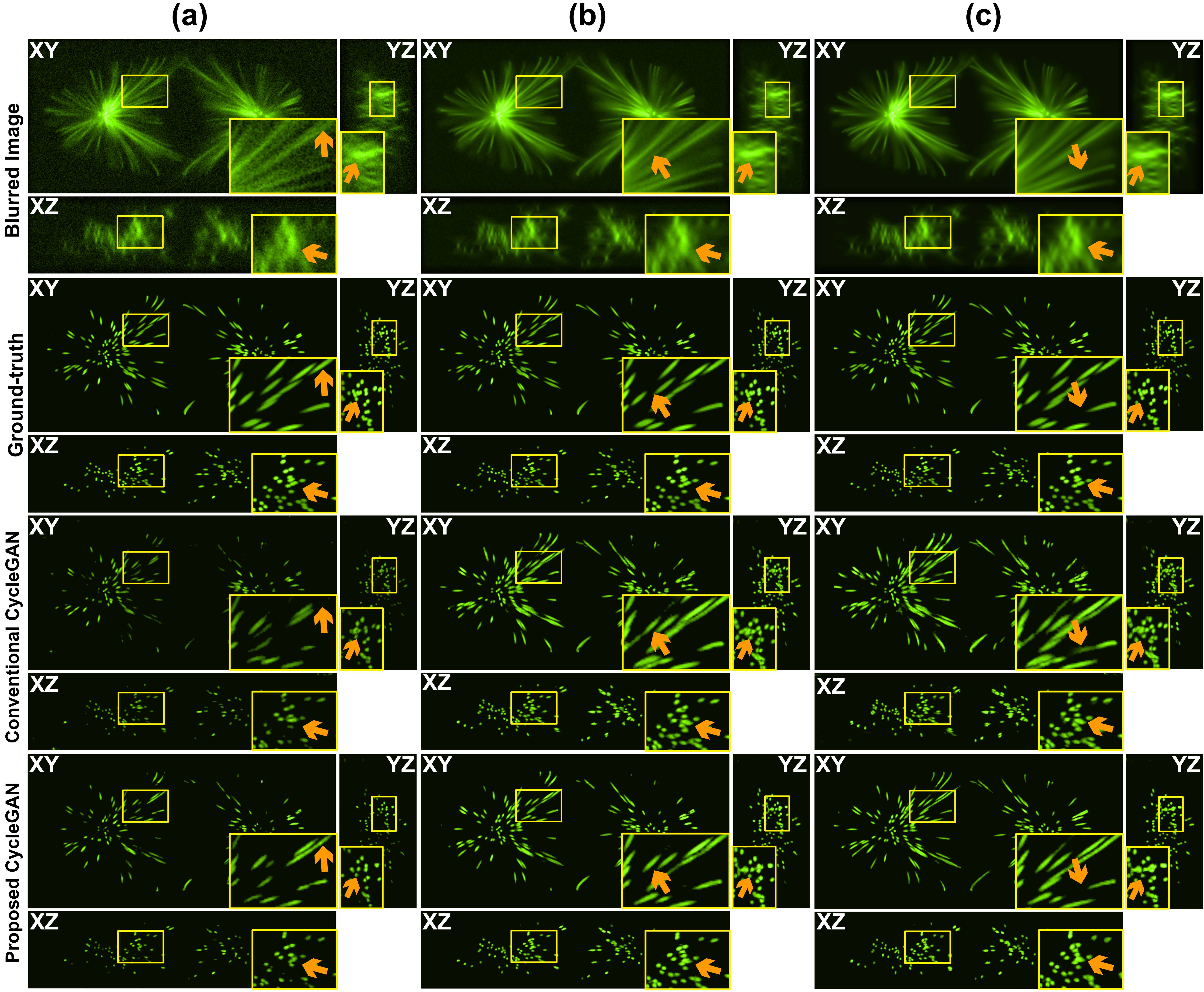}
\caption{Generalization across different levels of noise added after blurring: (a) \fix{average} SNR 10dB, (b) \fix{average} SNR 20dB: of Gaussian and Poisson noise \fix{(peak SNR=24dB)} was added after the blurring with a PSF, and (c) with no added noise. Our proposed model shows robustness across different added noise levels. The ROIs (marked yellow) show the area for the enlarged parts. \add{Orange arrows, as reference points, indicate the same location in each image.} XY, YZ, and XZ are 2D slicing planes of a 3D image volume.}
\label{fig:noise_test}
\end{figure*}

\subsubsection{Network training}
The size of the synthetic data was $256 \times 512 \times 128$, and 18 samples of the synthetic data were used for training and one each sample was for validation and testing, respectively. The pre-processing of data includes random cropping by $64\times64\times64$ pixels and random rotation by 90 degrees and random flipping by axis. The baseline model was trained on the blurred image volume with added noise that is measured to be 20dB in \fix{average} SNR \fix{and peak SNR of 24.08dB \cite{king2018}}. 

In the inference step, our testing of deconvolution by the baseline model is based on the following parameters: the noise level measured in \fix{average} SNR, and the PSF model. During the inference, input image volumes are cropped to $64\times 64\times 64$ pixel sub-volumes and then later merged back to a whole image volume with overlapping regions of $16\times 16\times 16$ pixels.
To update the optimizer of the network, we used the Adam optimizer  \cite{kingma2014adam} with the learning rate of 0.0001. In our simulation study, we set the cycle-consistency loss weight $\gamma$ (9) to be 10.0 and equal in both domain-transfer directions.

The proposed method was implemented in Python on  {Pytorch}, and GeForce GTX 1080 Ti GPUs were used for both training and validating/testing the network.

\subsection{Real Microscopy Experiments} % Epifluorescence data}
 We also used epifluorescence (EPF) data to validate our model {with real-world data}.  The samples for our real microscopy experiments were prepared as follows.
\subsubsection{Cell culture and immunocytochemistry }

Monkey kidney fibroblasts (COS-7 cell-line, ATCC) were cultured at $37^\circ C$ and 5\% CO2 in DMEM (Welgene) supplemented with 10\% FBS (Thermo Fisher Scientific). For immunocytochemistry, cells were fixed in 4\% formaldehyde, 4\% sucrose, PBS for 15 minutes, permeabilized for 5 min in 0.25\% Triton X-100, PBS and blocked for 30 min in 10\% BSA, PBS at $37^\circ C$. The cells were incubated with Tubulin antibody (AbCam), 3\% BSA, PBS for 2 hours at $37^\circ C$, washed in PBS, and incubated with Alexa-488 secondary antibody (Thermo Fisher Scientific), 3\% BSA, PBS for 45 min at $37^\circ C$.
\subsubsection{Epifluorescence Microscopy}
For three-dimensional epifluorescence microscopy, samples were mounted on a coverglass and imaged using N-SIM based on an inverted microscope (ECLIPSE Ti-E, NIKON), equipped with an oil immersion objective lens (Apo TIRF 100x, N.A. 1.49, NIKON) and an EMCCD camera (iXon DU-897, Andor Technology). The acquired datasets were comprised of 50 axial sections of 512 $\times$ 512 pixels. The voxel size of the reconstructed images was 64 nm in the x- and y-dimensions and 60 nm in the z-dimension, with 16-bit depth.

\subsubsection{Network training}
A total of 30 EPF 3D image samples of tubulin with a size of $512\times 512\times 51$ were used: 26 images for training, 1 for validation, and 3 for testing. As for unmatched high resolution reference data for our cycleGAN training,
 we used the EPF image volumes that were deblurred by using a commercial software AutoQuant X3 (Media Cybernetics, Rockville).

Image volumes both during the training and the inference were handled in the same procedure as for the simulation study: random cropping by $64\times64\times64$ pixels and random rotation by 90 degrees and random flipping by axis for pre-processing, and the same crop-and-merge technique for the inference phase. We also used the Adam optimizer with the learning rate of 0.001. We also set the cycle-consistency loss weight $\gamma$ to be 10.0 and equal in both domain-transfer directions.

\subsection{Comparison with other deep learning algorithms}
To validate the proposed method, we compared the deconvolution performance of our model with that of other deep-learning-based deconvolution algorithms: the conventional cycleGAN architecture \cite{lu2017conditional,lee2019three}, which uses a deep neural network for both generators, and a supervised learning network with Least Squared (LS) GAN Loss \cite{isola2017}, which is not cyclic, and a supervised learning network with MSE (Mean Squared Error) loss, which is implemented as a U-Net generator (Fig. \ref{fig:Unetwork}). For our deconvolution tasks, the supervised learning methods require the training data to be paired: blurred-deblurred. Finally, to test on the real-world data, we compared the deconvolution results with those by AutoQuant X3. 

%\sout{In AutoQuant X3 experiments, the PSF was estimated by the blind-deconvolution procedure in AutoQuantX3, thus it could aggravate the aberration issues in the z-axis.}   %this was addressed in the results. 

\subsection{Performance Metrics}

In the simulation study, the ground-truth data are available, so  we validated our model quantitatively using the peak signal-to-noise ratio (PSNR) and structural similarity index metric (SSIM) \cite{wang2004image}. The PSNR is defined as follows:
$$MSE = \frac{1}{N_xN_y}\sum_{i=0}^{N_x-1}\sum_{j=0}^{N_y-1}\Bigl[x(i,j)-\hat x(i,j)\Bigr]^{2}$$
$$PSNR = 10\log_{10}\Bigl(\frac{MAX_{x}^{2}}{MSE}\Bigr)$$
where $N_x$ and $N_y$ are $x$- and $y$- dimension of the ground-truth image $x$, 
$\hat x$ is the noisy image approximation, $MAX_{x}$ is the maximum possible pixel value of the image.
The SSIM is defined as follows:
$$SSIM(x,\hat x) = \frac{(2\mu_{x}\mu_{\hat x}+c_{1})(2\sigma_{x\hat x}+c_{2})}{(\mu_{x}^{2}+\mu_{\hat x}^{2}+c_{1})(\sigma_{x}^{2}+\sigma_{\hat x}^{2}+c_{2})}$$
where $\mu$ is the average of the image, $\sigma_x,\sigma_{\hat x}$ is the variance of the images $x$ and $\hat x$, $\sigma_{x\hat x}$ is the covariance of the images,
$c_{1} = (k_{1},L)^{2}$ and $c_{2}=(k_{2},L)^{2}$ are the variables to stabilize the division with weak denominator where $L$ is a dynamic range of the pixel intensities, and $k_{1} = 0.01$ and $k_{2} = 0.03$ by default.

\begin{figure}[!htb]
\begin{center}
\includegraphics[width=0.7\linewidth]{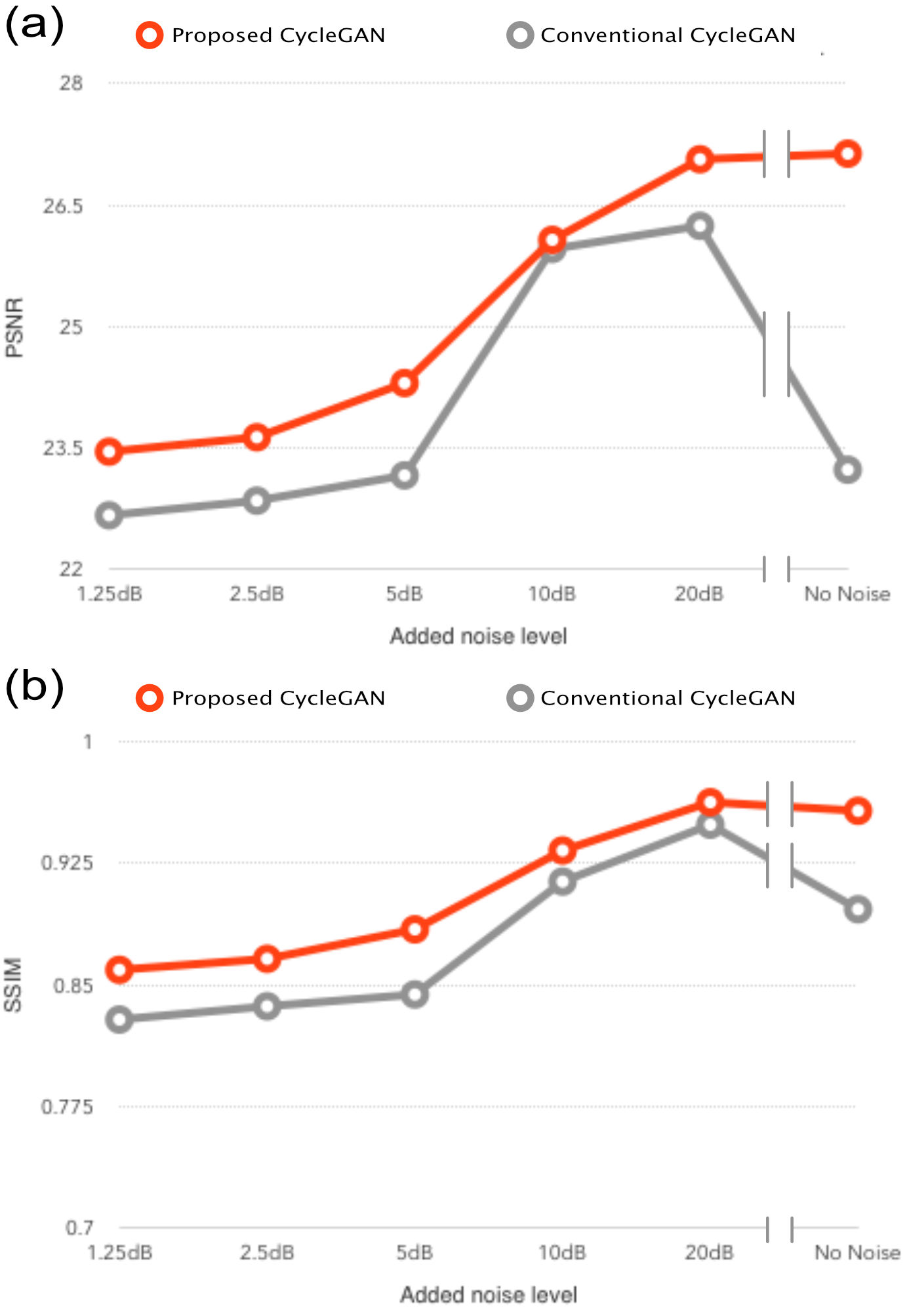}
\end{center}
\caption{Deconvolution performance in different levels of added noise after blurring. (a): PSNR (in dB), (b): SSIM}
\label{fig:noise_plot}
\end{figure}

\begin{figure}[htb!]
\centering\includegraphics[width=0.5\textwidth]{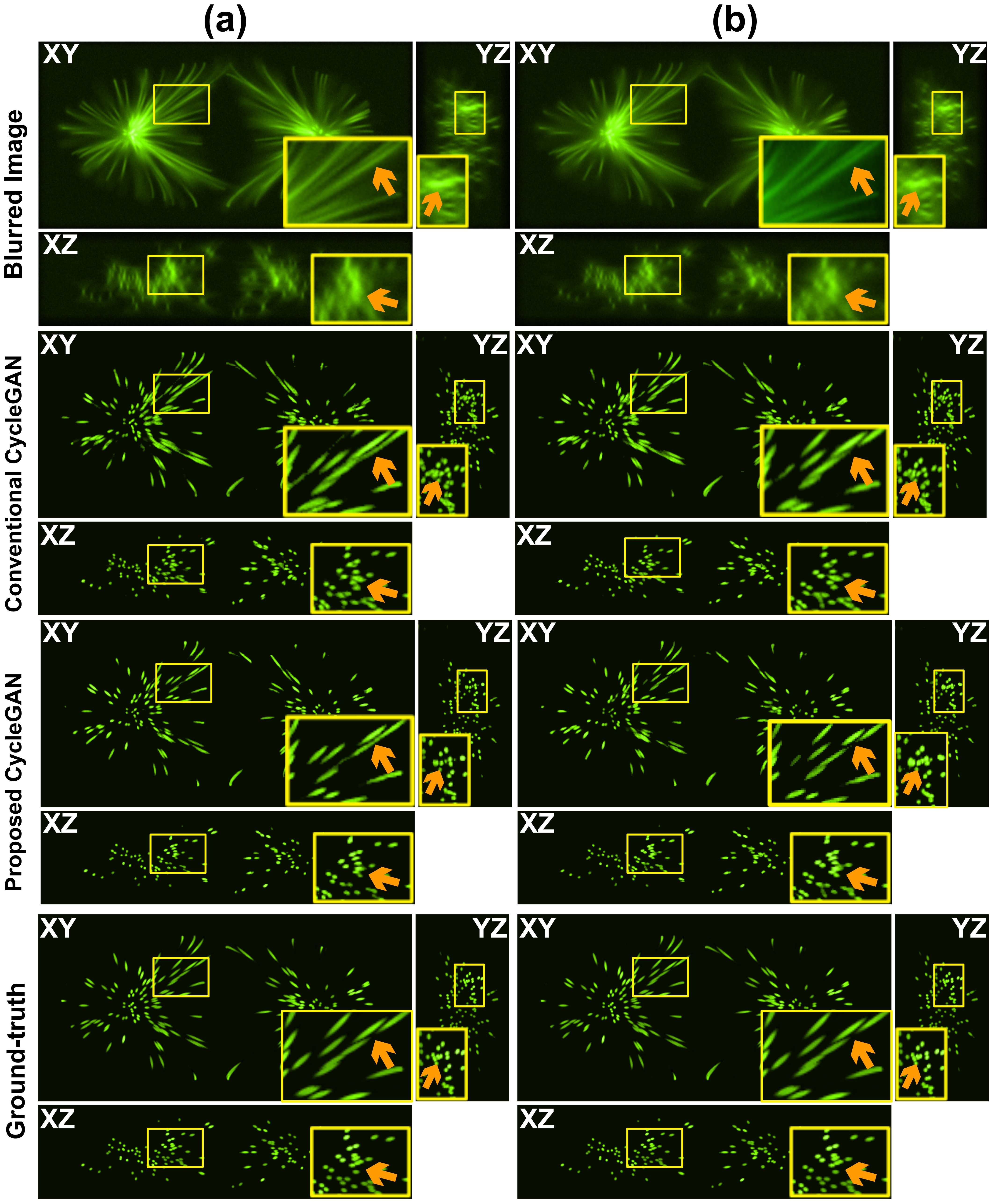}
\caption{Generalization across different PSF models: (a) Born {\&} Wolf model (b) Richards {\&} Wolf model. In both cases, \fix{average} SNR 20dB of Gaussian and Poisson noise \fix{(peak SNR= 24dB)} was added after the blurring with a PSF. The ROIs (marked yellow) show the area for the enlarged parts.  \add{Orange arrows, as reference points, indicate the same location in each image.} XY, YZ, and XZ are 2D slicing planes of a 3D image volume.}
\label{fig:psfs_test}
\end{figure}

\section{Experiments {\&} Results} 
\label{sec:results}

\subsection{Simulation Study}
In our simulation study, we first compare reconstruction performance of our proposed methods to the previous state-of-the-art deep learning algorithms. Our proposed methods are tested for robustness in two categories: added noise levels and PSF models. Furthermore, we visually interpret the PSF kernels estimated by our models. 

\subsubsection{Comparison with state-of-the-art Deep Learning algorithms}
Fig. \ref{fig:sota_comparison} and Table \ref{tab:sota_comparison} display the performance of our proposed model in comparison to the conventional cycleGAN model with the same LS-GAN loss, as well as to supervised methods that use either MSE Loss or LS-GAN loss. The supervised learning networks perform excellently for the deconvolution task; especially the supervised learning network with the MSE loss exhibits near state-of-the-art results. We observed the supervised learning network with GAN loss shows more cases where it suggests marginally better understanding of the tubule structure, while the one with MSE loss exhibits some false interlinkage of tubule objects (see Fig. \ref{fig:sota_comparison}).   In the cycleGAN models,
the proposed cycleGAN model shows the best results both in metrics and visuals (see Table \ref{tab:sota_comparison}). While we observed many cases of overestimation of the object region and loss of fine details from the conventional cycleGAN model, the proposed cycleGAN model shows far better understanding of the tubule structure.
In comparison with the supervised learning networks, the performance degradation in unsupervised learning
is unavoidable due to the lack of paired training data set.
Nonetheless, the proposed model provides visually similar reconstruction quality to the supervised learning approaches.

\begin{table}[!hbt]
\centering
\begin{tabular}{ccccc}
\hline
                                           &            & PSNR (dB)        & SSIM            \\ \hline\hline
\multicolumn{1}{c}{} 						 & Input      & 22.3020         & 0.7275          \\ %\cline{2-4} 
\multicolumn{1}{c}{}                     & Supervised learning with MSE loss & \textbf{30.232}   & \textbf{0.9826}         \\
\multicolumn{1}{c}{}                     & Supervised learning with GAN loss   & 29.0766          & 0.9774          \\ \hline
\multicolumn{1}{c}{}                     & Conventional cycleGAN    & 26.2483          & 0.9474         \\  
\multicolumn{1}{c}{}                     & Proposed cycleGAN    &\textbf{ 26.6249}        & \textbf{0.9573}   \\  
\hline
\end{tabular}
\caption{Deconvolution performance in comparison with state-of-the-art deep learning algorithms \add{(refer to Fig. \ref{fig:sota_comparison} )} in terms of PSNR and SSIM. }
\label{tab:sota_comparison}
\end{table}

\subsubsection{Generalization with respect to added noise levels}
We tested performance of our proposed model and the conventional cycleGAN model with respect to higher levels of added noise: 1.25dB, 2.5dB, 5dB, 10dB, and also no added noise. The mixture of Gaussian and Poisson noise was added after the PSF convolution, and the models were trained with simulated images with added noise of \fix{average} SNR 20dB \fix{with peak SNR 24dB}. Fig. \ref{fig:noise_test} and Fig. \ref{fig:noise_plot} display such results. While performance of either model deteriorates as the added noise level increases, our proposed model maintains consistent superiority in both visuals and metrics.  Furthermore, our model is still robust when no noise was added. In contrast, the performance metrics of the conventional cycleGAN model, especially its PSNR, decline sharply when there is no added noise. We speculate that this decline is due to the conventional model's heavily trained denoising feature, whose learning can more easily be over-fitted by the deeper generator network in the blurring path, and which thereby leads to overestimation of the object regions and pixel intensities when there is no added noise present.

\subsubsection{Generalization with respect to PSF models}
We compared the baseline performance of our proposed model and the conventional cycleGAN model to that of the models on the images blurred by an alternative PSF model: the Richards {\&} Wolf model. Fig. \ref{fig:psfs_test} and Table \ref{tab:psfs_test} show these results. The same level of noise (\fix{average} SNR 20dB \fix{with peak SNR 24dB}) was added after the blurring. In testing for both PSF models, the proposed cycleGAN model performed the best both in visuals and metrics. 

The Richards {\&}  Wolf model describes the vectorial-based diffraction instead of the scalar-based diffraction. Despite this fundamental difference, all the models show comparable performances for the Richards {\& } Wolf model.

\begin{table}[!hbt]
\centering
\begin{tabular}{ccccc}
\hline
                                           &            & PSNR (dB)        & SSIM            \\ \hline\hline
\multicolumn{1}{c}{}  {\multirow{4}{*}{(a)}}	 & Input      & 22.3020       & 0.7275          \\ 
\multicolumn{1}{c}{}                     & Conventional cycleGAN   & 26.2483          & 0.9474         \\  
\multicolumn{1}{c}{}                     & Proposed cycleGAN    &\textbf{ 26.6249}    & \textbf{0.9573}   \\ \hline
\multicolumn{1}{c}{}  {\multirow{4}{*}{(b)}}	 & Input      & 22.5202       & 0.7348         \\ 
\multicolumn{1}{c}{}                     & Conventional cycleGAN & 27.2886          & 0.9584         \\  
\multicolumn{1}{c}{}                     & Proposed cycleGAN  		& \textbf{27.5286}     & \textbf{0.9653}   \\  \hline

\end{tabular}
\caption{Generalization across different PSF models. (a) Born {\&} Wolf model, (b) Richards{\&} Wolf model. See Fig. \ref{fig:psfs_test}}
\label{tab:psfs_test}
\end{table}

\subsubsection{Comparison with non-blind deconvolution}
\begin{figure}[htb!]
\centering\includegraphics[width=0.5\textwidth]{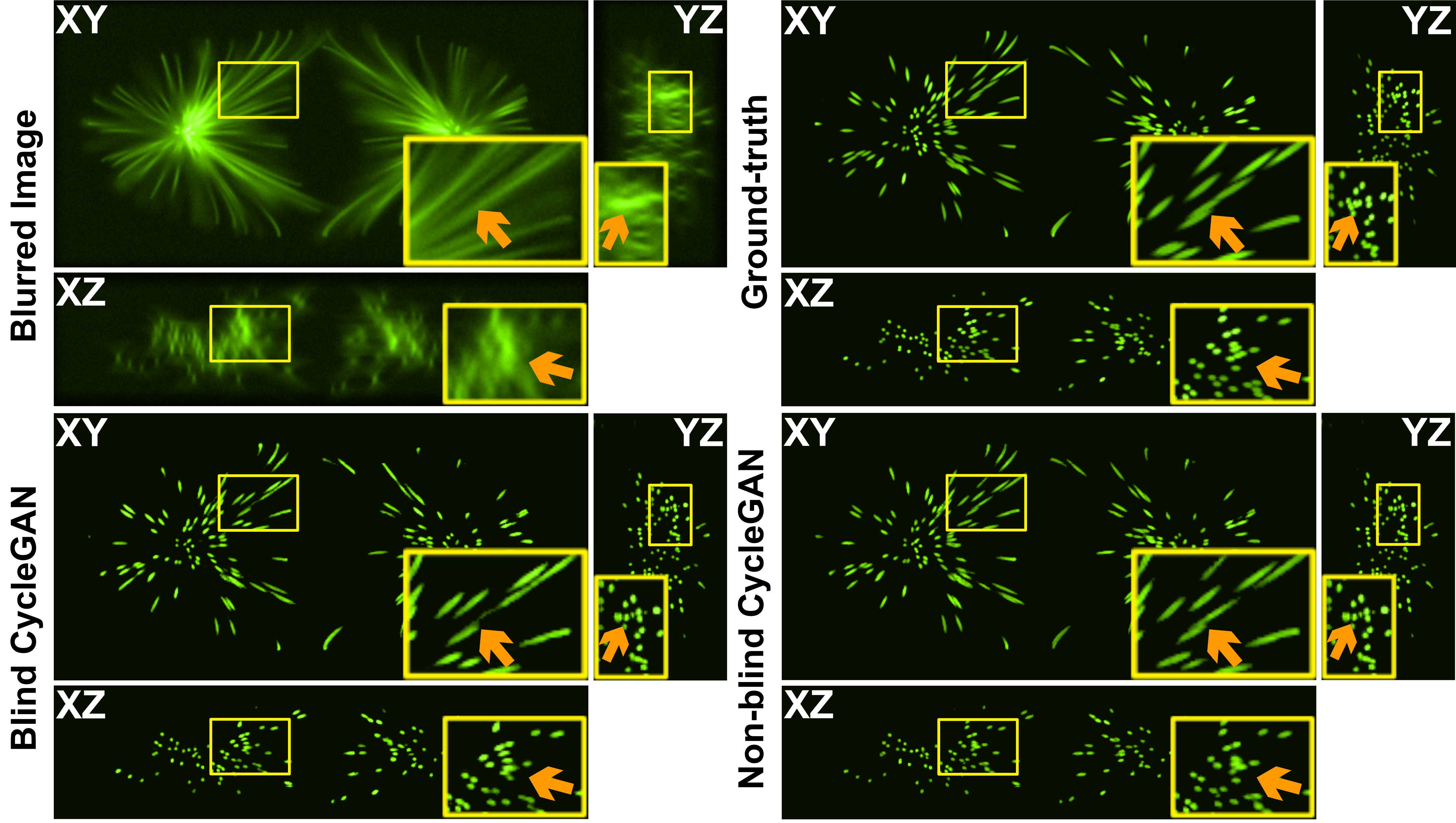}
\caption{Comparison with the non-blind deconvolution scenarios by our proposed cycleGAN model. Both blind cycleGAN model and non-blind cycleGAN models refer to models based on our proposed cycleGAN architecture. The ROIs (marked yellow) show the area for the enlarged parts. \add{Orange arrows, as reference points, indicate the same location in each image.} XY, YZ, and XZ are 2D slicing planes of a 3D image volume.}
\label{fig:blind_test}
\end{figure}

We evaluated the proposed cycleGAN approaches in a non-blind deconvolution set-up, where the blur kernel $h(\rb)$ is known. In our implementation, the blur kernels were replaced with the same PSF kernels that were used to generate the original blurred images, and the corresponding discriminator was removed as shown in Fig.~\ref{fig:cycleGAN}(b) since the fixed kernels do not learn. The non-blind deconvolution setting is designed to reveal the reconstruction capability of the network without its need to estimate the PSF during training. 
Fig. \ref{fig:blind_test} and Table \ref{tab:blind_test} show the results. 
As aligned with our initial assumption, the non-blind variations of our cycleGAN model show considerable improvement for reconstruction both in metrics and visuals.

\begin{table}[!hbt]
\centering
\begin{tabular}{ccccc}
\hline
                                           &            & PSNR (dB)        & SSIM            \\ \hline\hline
 & Input      & 22.3020       & 0.7275          \\ %\cline{2-4} 
\multicolumn{1}{c}{}                     & Proposed cycleGAN  (blind case)   &26.6249   & 0.9573 \\
\multicolumn{1}{c}{}                     & Proposed cycleGAN  (non-blind)  &\textbf{ 29.2925}    & \textbf{0.9746} \\ \hline
\end{tabular}
\caption{Comparison with the non-blind deconvolution scenarios by our proposed cycleGAN model. See Fig. \ref{fig:blind_test}}
\label{tab:blind_test}
\end{table}

\subsubsection{Estimation of the learned PSF kernels}
\begin{figure}[!htb]
\begin{center}
\includegraphics[width=0.9\linewidth]{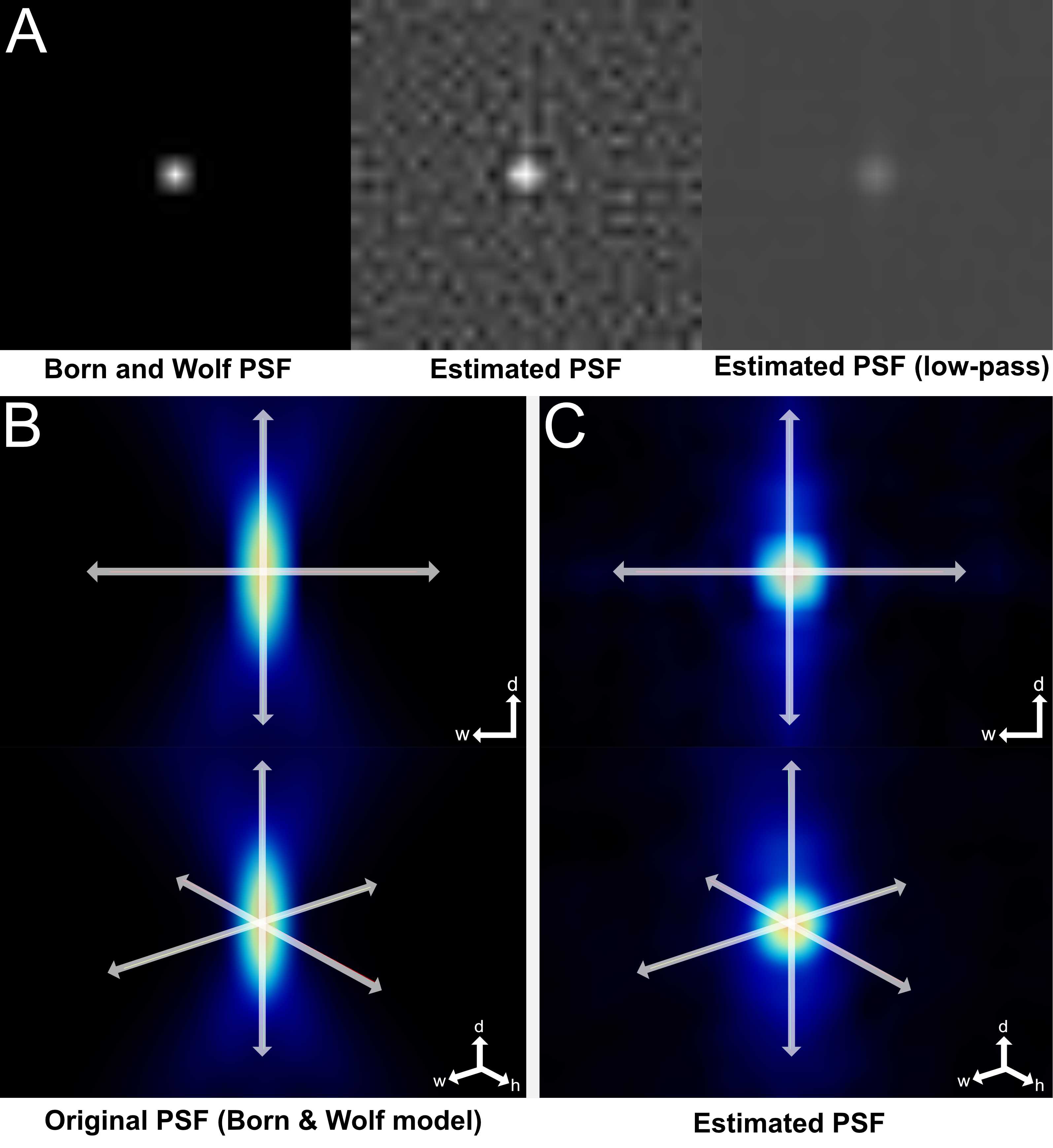}
\end{center}
\caption{Examination of the estimated PSF kernels: (A) 2D images of PSF kernels at the center: note that the raw reconstructions include much high frequency data. (B)(C) \add{3D visualizations in two different views. (B) The original PSF kernel (Born and Wolf), (C) the estimated PSF kernel by the proposed cycleGAN} 
 (d, h, w) are  depth, height, width of an image volume. }
 
\label{fig:estimatedpsfs}
\end{figure}

In order to further examine the learning of deconvolutional features by our proposed method, we reconstructed a PSF kernel that is learned after the training. Fig. \ref{fig:estimatedpsfs} shows its visualizations. The raw reconstructions exhibit  remote resemblances to the original PSF kernel, containing high frequency features: see Fig. \ref{fig:estimatedpsfs}(a). We assume the prevalence of such high frequency data is due to simulating the added noise condition from the original blurring process. In order to recover the underlying PSF kernel structure, we  applied 3D Gaussian blurring ($\sigma = 1$) to filter out the high frequency data. Then, we could observe the estimated PSF kernel structures: see Fig.\ref{fig:estimatedpsfs}(b)(c). 
It is now clear that they do form apparent appearances of a PSF kernel with respect to their impulse response. Therefore, we regard this as evidence that our proposed networks do encourage their linear kernels to emulate the blurring process. This observation also suggests the possibility of our proposed methods to be considered a more interpretable form of a deep learning based deconvolution module. 

Although the estimated PSF in microscopy still has some common visual features of a PSF in terms of general elongation (visible as faded blue) in the longitudinal axis, as opposed to its absence in the lateral axis, we also observed that
there exists a discrepancy, especially along the axial direction. We believe that the main reason for this is that the linear layer not only learns the PSF but also noise generation processes. This is the current limitation of this work, and we believe that decoupling the noise generation processing and the PSF layer could overcome this limitation. This is an interesting topic for future research.

\subsection{Real Epifluorescence Microscopy Data}

\begin{figure*}[h!]
\centering\includegraphics[width=0.9\textwidth]{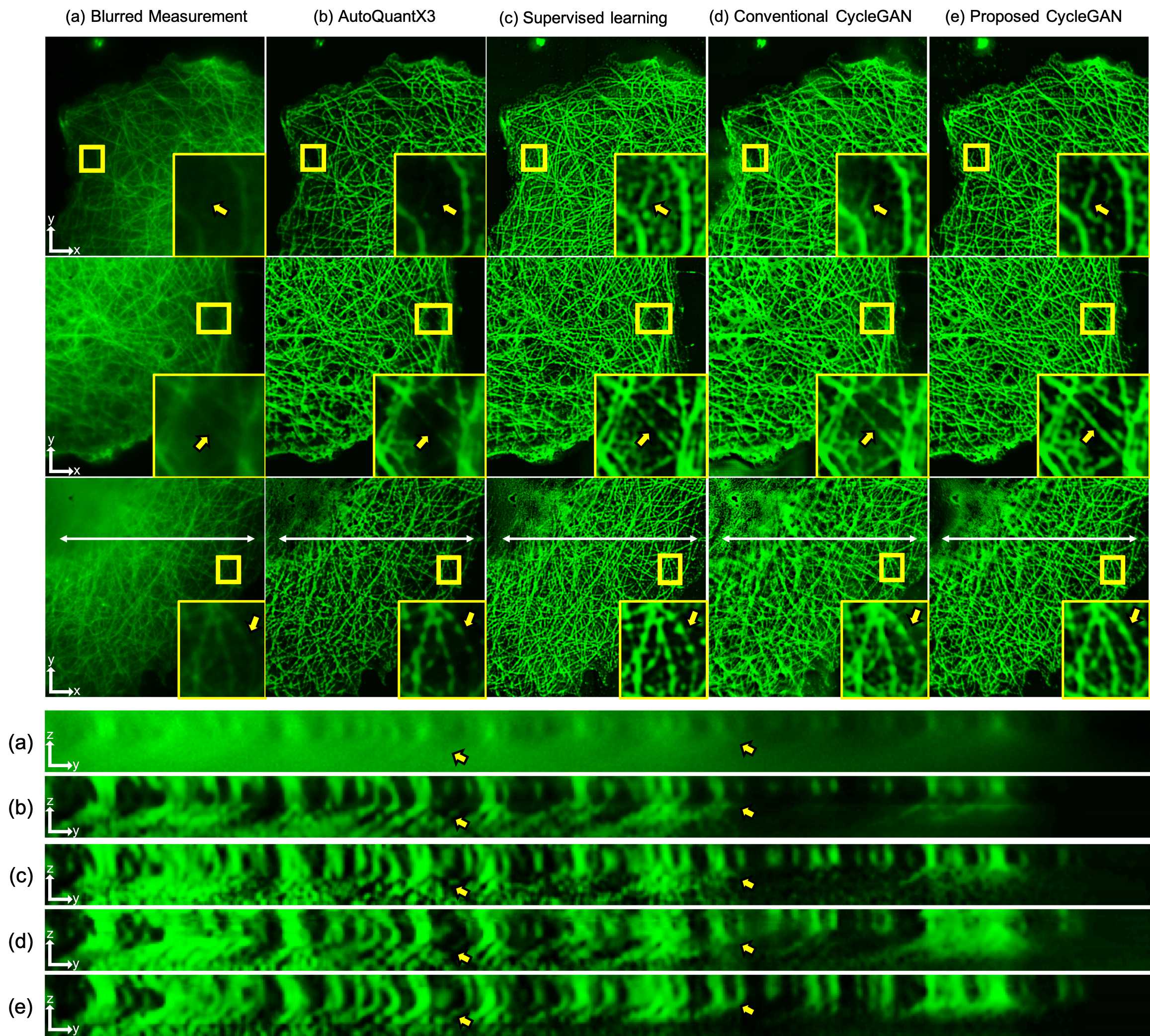}
\caption{Comparison of reconstruction results by various methods: (a) Blurred image measurements, (b) AutoQuantX3, (c) supervised learning, (d) the conventional cycleGAN,
and (e) the proposed cycleGAN.  The ROIs (marked yellow) show the area
for the enlarged parts. \add{Yellow arrows, as reference points, indicate image regions where different structures are more clearly observed. X-Y views are in the top panel and Y-Z views are in the bottom panel. }}
\label{fig:cell}
\end{figure*}

Fig. \ref{fig:cell} shows the results of application on the actual bio-imaging data. Since the original data about the PSF convolution is unavailable, there is no real ground-truth for this deconvolution task. Although one may want to use AutoQuant X3,  a widely used deconvolution software in practice, as the ground-truth, care should be taken.  Since the classical iterative blind deconvolution methods such as AutoQuant X3 is based on the top-down model assumption such as sparsity, smoothness, etc of image and the PSF, if these assumptions do not hold, there are chances that the resulting deconvolution results are not correct. In fact, AutoQuant X3 is known to generate inaccurate depth resolution at times, thereby leaving image regions at certain depths still partially blurry.
In our experiment, we also observed that while AutoQuant X3 exhibits some level of deconvolution, the model leaves out fine details of some tubule structures. %\sout{with some regions still remaining blurry.}

The supervised learning networks display more clarity in details and contrast, revealing some covered structures that were previously invisible in the input images. However, the surrounding regions seem overexposed and do not clarify the salient geometry of the micro-structures, as evidenced by some tubule structures remaining disconnected.

The proposed cycleGAN model with LS-GAN loss shows the most noticeable improvement in image quality, which is well illustrated by revealing the micro-structures, especially in the second image sample in the Fig. \ref{fig:cell}. We cross-checked the presence of such micro-structures from their blurred image counterparts in the neighboring z-stack slices, where aberration is less severe and the corresponding structures were visible in a coherent manner, and they were also deemed biologically plausible by our in-house experimental biologists. 
%In our interpretation,  such newly recovered features by the proposed method are not artificial but originate from blurred measurements,
As our network is trained to learn not from direct correlation of the data pairs but  to minimize the
{\em average} transport cost,  the trained network is observed to generalize better,  and
 more likely to learn more from various image regions. Similar advantages were observed in other imaging applications
 \cite{sim2019optimal}. Therefore,  we believe that such newly recovered features by the proposed deconvolution
  method are not artificial but recovered from blurred measurements.

The conventional cycleGAN model also recovers these micro-structures to some extent but does not exhibit the same level of clarity.

\subsection{Reduction of Computational Cost}
In our implementations, while a conventional cycleGAN model has on average 7 million parameters for its generator in the blurring kernels, our proposed cycleGAN model has on average 0.03 million parameters for its linear blur kernel, thereby reducing the memory size of the parameters for training to 61.3$\%$ of the conventional cycleGAN.

The training time for the proposed network was 10 hours for the simulation study, and 18 hours for the real microscopy experiments, respectively. At inference phase, the computational time for $256\times 512\times 128$ volume image in the simulation study was {21 sec}, and it was 15 sec for $512\times 512\times 50$ size  volume for the real microscopy experiments.

\section{Conclusions}
\label{sec:conclusion}

In this paper, we presented a novel cycleGAN architecture with a linear blur kernel
for deconvolution microscopy, which can be used for both non-blind and blind deconvolution problems in an unsupervised manner. In contrast to the existing cycleGAN approaches with two CNN-based generators,
in the proposed method only a single CNN-based generator is necessary to convert the low resolution image to high resolution one, whereas
the blur image is generated using a simple linear layer.
For the case of non-blind setup, this structure can be further simplified to a single pair of generator and discriminator.
We showed that the proposed architecture is indeed a dual formulation of an optimal transport problem that uses
a novel  PLS cost as the transport cost  between two measures  for low- and high-resolution images.

In contrast to the standard cycleGAN with two deep generators, the proposed
architecture uses the physics-driven convolution model as a generator so that
the data consistency can be imposed automatically during the training.
Furthermore, due to the fewer number of the unknown parameters,
the network training is not only more stable but also cost-efficient compared to the standard cycleGAN with two deep generators. We also showed in our experiments that despite its lowered computational cost, our method does not sacrifice its reconstruction capabilities in learning non-deconvolutional features compared to the conventional cycleGAN.
% as well. 
Since our design encourages the linear kernel to simulate the PSF-convolution process, we also explored the possibility of having a more interpretable deep learning module as we can explicitly examine the estimated PSF kernel after the training.

 We used the synthetic microtubule network and the real epi-fluorescent  microscopy data to validate our model.  In both non-blind and blind setups, the proposed method accurately reconstructed high resolution images. 
 Moreover, our experimental results show that our method  generalized well for the deconvolution
 problems in different PSF models as well as noise conditions.
Given the robustness and lowered computational cost, we believe that our method can be an important platform for deconvolution microscopy.

\section*{Acknowledgment}

This work is supported by  National Research Foundation of Korea (Grant NRF-2020R1A2B5B03001980 and NRF-2017M3C7A1047904).

\bibliographystyle{IEEEtran}
\bibliography{refs.bib}

\end{document}

%% file: packages.tex
\usepackage{outline}
\usepackage{pmgraph}
\usepackage[normalem]{ulem}
\usepackage[utf8]{inputenc}
\usepackage{amssymb}
\usepackage{hyperref}
\usepackage{amsmath}
\usepackage{graphicx}
\usepackage{times}
\usepackage{xcolor}
\usepackage{xspace}
\usepackage[colorinlistoftodos]{todonotes} % for \todo
\usepackage{cite}
\usepackage{bm} % bold
\usepackage{url}

\usepackage{epstopdf}
\AppendGraphicsExtensions{.tiff}
\graphicspath{{fig/}} % Directory in which figures are stored

\usepackage{epsfig}
\usepackage{tikz}
\usetikzlibrary{spy}
\usepackage{algpseudocode}
\usepackage{algorithm}
\usepackage{mathrsfs}

\usepackage{amsmath,graphicx,caption,mathtools,amssymb,amsthm}
\usepackage{graphicx}
\usepackage{multirow}
\usepackage{subcaption}
\usepackage{accents}

%% file: macros.tex
\newcommand{\add}[1] {\textcolor{blue}{#1}} % for revision

\def\QED{~\rule[-1pt]{5pt}{5pt}\par\medskip}

\long\def\comment#1{} % comment out text

\newcommand{\xmath}[1] {\ensuremath{#1}\xspace}
\newcommand{\blmath}[1] {\xmath{\bm{#1}}}

\newcommand{\rb}{{\blmath r}}

\newcommand{\Rc}{\mathcal{R}}

\newcommand{\Hc}{\mathcal{H}}

\newcommand{\Nc}{\mathcal{N}}

\newcommand{\Xc}{\mathcal{X}}
\newcommand{\Yc}{\mathcal{Y}}

\newcommand{\Rd}{{\mathbb R}}

\newcommand{\Kd}{\mathbb{K}}

\newcommand{\beq}{\begin{equation}}
\newcommand{\eeq}{\end{equation}}
\newcommand{\beqa}{\begin{eqnarray}}
\newcommand{\eeqa}{\end{eqnarray}}